# Second-Harmonic Generation in Etchless Lithium Niobate Nanophotonic Waveguides with Bound States in the Continuum

Fan Ye, Yue Yu, Xiang Xi, and Xiankai Sun*

*Department of Electronic Engineering, The Chinese University of Hong Kong, Shatin, New Territories, Hong Kong*

**Corresponding author:* xksun@cuhk.edu.hk

**Abstract:** Bound states in the continuum (BICs) have been extensively studied in various systems since its first proposal in quantum mechanics. Photonic BICs can enable optical mode confinement and provide field enhancement for nonlinear optics, but they have rarely been explored in nonlinear integrated photonic waveguides. Applying BICs in photonic integrated circuits enables low-loss light guidance and routing in low-refractive-index waveguides on high-refractive-index substrates, which is suitable for integrated photonics with nonlinear materials. Here, we report experimental demonstration of second-harmonic generation from telecom to near-visible wavelengths on an etchless lithium niobate platform by using a photonic BIC for the second-harmonic mode. The devices feature second-harmonic conversion efficiency of 0.175% $W^{-1}$ $cm^{-2}$ and excellent thermal stability with a wavelength shift of only 1.7 nm from 25°C to 100°C. Our results represent a new paradigm of nonlinear integrated photonics on a cost-effective and convenient platform, which can enable a broad range of on-chip applications such as optical parametric generation, signal processing, and quantum photonics.

**Keywords:** *lithium niobate, integrated photonics, etchless process, second-harmonic generation, bound states in the continuum*

## Introduction

The concept of bound states in the continuum (BICs) was first proposed by von Neumann and Wigner[1] in 1929 with mathematical construction of a three-dimensional potential that can support perfectly confined states in a continuous band. The radiation loss of these confined states can be eliminated by destructive interference with the continuous modes. Such a phenomenon has been found in many physical systems that exhibit the properties of waves, like electronics[2-5], acoustics[6-9], and photonics[10-20]. Photonic BICs were explored in photonic crystals[11, 20, 21], optical gratings[12], and metasurfaces[15, 17], rendering new applications in lasers[18], sensors[17, 21], and filters[12]. Recently, it was proposed and demonstrated that photonic BICs can exist in a low-refractive-index waveguide on a high-refractive-index substrate[22, 23]. Confined transversely to the region of high-refractive-index substrate below the low-refractive-index waveguide, light flows along the low-refractive-index waveguide under its longitudinal guidance. The destructive interference among various loss channels leads to forbidden energy dissipation in the BIC-based waveguide, resulting in theoretically zero propagation loss[22]. Based on that, photonic integrated circuits including low-loss waveguides and high-$Q$ microcavities were obtained without the need for etching[23]. To date, the applications of this type of photonic BICs are in the realm of linear optics. Their applications in nonlinear optics are yet to be demonstrated.

The second-order ($\chi^{(2)}$) nonlinear optical processes[24] include second-harmonic generation, sum/difference-frequency generation, and parametric down-conversion, which play important roles in wavelength conversion[25-28], photon-pair generation[29, 30], and supercontinuum generation[31, 32]. As the "silicon of photonics," lithium niobate is arguably the most popular nonlinear optical material[30, 32-40] because of its large $\chi^{(2)}$ nonlinear coefficients and wide transparency window (350 nm to 5 μm). To date, nonlinear optical processes have been realized on integrated lithium niobate platforms by lithographic patterning and dry etching of the lithium niobate thin film[28, 30, 32, 35, 41]. The performance of the fabricated devices relies crucially on the quality of the dry etching process. However, obtaining etched structures in lithium niobate with high quality is not trivial. The fabrication requires expensive equipment and substantial effort in process development.

Here, we report experimental demonstration of efficient second-harmonic generation on a lithium niobate integrated platform which does not need etching of lithium niobate. By fabricating a polymer waveguide with carefully chosen dimensions on a lithium-niobate-on-insulator substrate, we obtained modal phase matching between the orthogonally polarized fundamental and second-harmonic modes, both with low propagation loss, at the telecom and near-visible wavelengths respectively, where the second-harmonic mode is a TM-polarized BIC. With lateral confinement by the waveguide structure, the fundamental mode and the second-harmonic mode are colocalized in the lithium niobate layer with a large modal overlap. The measured maximal second-harmonic generation efficiency from a 5-mm-long waveguide is 0.175% $W^{-1}$ $cm^{-2}$. The thermo-optic coefficients of the polymer and lithium niobate have opposite signs, which results in excellent thermal stability with a wavelength shift of only 1.7 nm from 25°C to 100°C. As the first experimental demonstration of nonlinear optics with photonic BICs on a chip, our devices have set a new paradigm for exploring optical nonlinearities and light–matter interactions on an etchless integrated photonic platform, with the advantages of simplified fabrication processes and improved thermal stability.

**Designing BIC waveguide for second-harmonic generation**

Figure 1a illustrates the cross-sectional structure of the device for second-harmonic generation. The devices were fabricated on a lithium-niobate-on-insulator (LNOI) substrate, with 150-nm-thick *z*-cut lithium niobate layer (blue) on 2-µm-thick silicon oxide (gray). We patterned a polymer waveguide (green) with width $w$ and thickness $t$ on the substrate. This waveguide structure supports multiple modes with different polarizations. We chose the fundamental transverse-electric ($TE_{00}$) mode for the pump light and the second-order transverse-magnetic ($TM_{20}$) mode for the second-harmonic light, which can satisfy the modal phase-matching condition that is required for second-harmonic generation. Figures 1b and 1c depict the electric field ($|\mathbf{E}|$) profiles of the $TE_0$ mode at the pump wavelength $\lambda_{pump}$ = 1559.4 nm and of the $TM_{20}$ mode at the second-harmonic wavelength, respectively, simulated with a finite-element method. It is important to obtain low propagation loss in both the $TE_{00}$ and $TM_{20}$ modes to achieve high efficiency for the second-harmonic generation.

For second-harmonic generation in a waveguide with a length $L$, the generated second-harmonic power $P_{SHG}$ under the nondepleted approximation can be expressed as[34]

$$P_{\mathrm{SHG}} = \eta P_{\mathrm{pump}}^2 g(L). \tag{1}$$

In Eq. (1), $P_{pump}$ is the pump power. $\eta$ is the normalized second-harmonic conversion efficiency expressed as

$$\eta = \frac{8\pi^2}{\varepsilon_0 c n_{\omega}^2 n_{2\omega} \lambda_{\omega}^2} \frac{\xi^2 d_{\text{eff}}^2}{A_{\text{eff}}}, \tag{2}$$

where $\varepsilon_0$ and $c$ are respectively the permittivity and light speed in vacuum, $d_{\text{eff}}$ is the effective nonlinear susceptibility, $\xi$ is the spatial modal overlap between the fundamental and second-harmonic modes. $A_{\text{eff}} \equiv (A_1^2 A_2)^{1/3}$ is the effective modal area with

$$A_i = \frac{\left(\int_{\text{all}} |\mathbf{E}_i|^2 \, dxdz\right)^3}{\left|\int_{\chi^{(2)}} |\mathbf{E}_i|^2 \, \mathbf{E}_i dxdz\right|^2}, \ (i = 1, 2) \tag{3}$$

where $\int_{\chi^{(2)}}$ and $\int_{\text{all}}$ denote two-dimensional integration over the lithium niobate thin film (silicon oxide and polymer do not have second-order nonlinearities) and over all space, respectively. $\mathbf{E}_1$ and $\mathbf{E}_2$ are the electric fields of the fundamental and second-harmonic mode, respectively. $g(L)$ is the loss factor defined as

$$g(L) = e^{-(\alpha_{\omega}+\alpha_{2\omega})L} \frac{(e^{\Delta\alpha L}+1)^2 - 4\cos^2(L\Delta k/2)e^{\Delta\alpha L}}{(\Delta\alpha)^2 + (\Delta k)^2}, \tag{4}$$

with $\Delta\alpha = \alpha_{\omega} - \alpha_{2\omega}/2$, where $\alpha_{\omega}$ and $\alpha_{2\omega}$ are the attenuation coefficients of the fundamental and second-harmonic mode, respectively. $\Delta k$ is the phase mismatch factor defined as $4\pi(n_{2\omega} - n_{\omega})/\lambda$ where $n_{2\omega}$ and $n_{\omega}$ are the effective refractive indices of the TE mode at the fundamental wavelength and the $TM_{20}$ mode at the second-harmonic wavelength, respectively. Equation (1) shows that the generated second-harmonic power depends not only on the theoretical normalized conversion efficiency $\eta$ but also on the loss factor $g(L)$ of the waveguide. Under the assumption of zero propagation loss, $P_{\text{SHG}}$ can be simplified as

$$P_{\text{SHG}} = \frac{8\pi^2}{\varepsilon_0 c n_{\omega}^2 n_{2\omega} \lambda_{\omega}^2} \frac{\xi^2 d_{\text{eff}}^2}{A_{\text{eff}}} P_{\text{pump}}^2 L^2 \operatorname{sinc}^2(L\Delta k/2). \tag{5}$$

In this waveguide structure, the $TE_{00}$ mode as a regular bound mode always has negligible propagation loss. However, since the TE modes have a higher effective refractive index than the TM modes, the $TM_{20}$ bound mode lies in the continuum of the TE modes. Light in the $TM_{20}$ mode can interact with the TE continuous modes, causing energy dissipation to the substrate and introducing nonnegligible propagation loss. The energy is dissipated to the TE continuous modes at the two edges of the waveguide through multiple channels. The decay length $L$ of the $TM_{20}$ bound mode propagating in the waveguide can be expressed as $L \propto w^2/\sin^2(k_y w/2)$, where $k_y$ is the $y$ component of the wave vector of the TE continuous mode which matches that of the $TM_{20}$ bound mode and $w$ is the width of the waveguide[22]. When $k_y w$ is a multiple of $2\pi$, $L$ approaches infinity and the $TM_{20}$ bound mode becomes a BIC correspondingly. Figure 1d shows the simulated propagation loss of the $TM_{20}$ mode in a straight waveguide as a function of the waveguide width $w$ and wavelength $\lambda$, with the waveguide thickness $t$ fixed at 600 nm. In the simulated wavelength range, the BIC can be obtained when the waveguide width $w$ is near 1.5, 2.5, and 3.4 μm. On the other hand, we also need to ensure modal phase matching between the $TE_{00}$ and $TM_{20}$ modes for efficient second-harmonic generation, which means that the effective refractive index of the $TE_{00}$ mode at the fundamental wavelength must match that of the $TM_{20}$ mode at the second-harmonic wavelength.

Therefore, for a LNOI wafer with a given thickness of lithium niobate layer, we designed the optimal thickness $t$ and width $w$ of the polymer waveguide to achieve the BIC and phase-matching conditions. For a given waveguide thickness $t$, we simulated the propagation loss and effective refractive index for the $TE_{00}$ mode at the fundamental wavelength and the $TM_{20}$ modes at the second-harmonic wavelength for the structure in Fig. 1a with a varying waveguide width $w$. The BIC point is where the $TM_{20}$ mode has zero propagation loss and the phase-matching point is where the effective refractive indices for the two modes are equal, as shown in Figs. 1f and 1g. The required waveguide widths for achieving the BIC and phase-matching conditions as a function of the waveguide thickness $t$ are then plotted in Fig. 1e. The intersection point of the two curves show that the optimal design parameters are $t$ = 600 nm and $w$ = 2.48 μm, with which the BIC and phase-matching conditions can be achieved simultaneously. For waveguides of this optimal design, we obtained a theoretical normalized second-harmonic conversion efficiency $\eta$ of 0.56% $W^{-1}$ $cm^{-2}$, where $d_{eff} = d_{31}$ = 4.3 pm/V was adopted in the calculation for the TE–TM conversion in $z$-cut lithium niobate.

It should be noted that the second-harmonic conversion efficiency depends mainly on the scheme of phase matching. Our devices use the scheme of modal phase matching without the need for periodic poling (much simpler device fabrication), while most of the traditional waveguides exhibiting higher conversion efficiency use quasi-phase matching by periodic poling (much more complicated device fabrication). By using the scheme of modal phase matching, we need to select a fundamental TE mode for the fundamental wavelength and a high-order TM mode for the second-harmonic wavelength, which results in a small modal overlap. By using the scheme of quasi-phase matching, there is no such restriction and thus the fundamental TE mode can be used for both fundamental and second-harmonic wavelengths, achieving the maximal modal overlap. Therefore, the conversion efficiencies of periodically poled lithium niobate waveguides are usually 2–3 orders of magnitude higher than those without periodic poling (See Tables S1 and S2 in the Supporting Information). If we were to use periodically poled lithium niobate in combination with the BIC-based waveguides for second-harmonic generation, the theoretical conversion efficiency can achieve 2571% $W^{-1}$ $cm^{-2}$, i.e., a factor of ~5000 enhancement upon our current scheme (see Section S2 in the Supporting Information).

**Device fabrication and characterization**

We fabricated the devices with an etchless process on a LNOI wafer supplied by NANOLN, with 150-nm-thick *z*-cut lithium niobate layer on 2-μm-thick silicon oxide on a silicon substrate handle. We spun-coated a 600-nm-thick polymer (ZEP520A) layer on the LNOI wafer with a spinning speed of 2000 r/min. Then, we defined the patterns in the polymer with high-resolution electron-beam lithography. As shown in Fig. 2a, an entire device consists of a long main waveguide for second-harmonic generation, grating couplers for coupling light between the on-chip waveguides and optical fibers (Fig. 2c), and directional couplers (Fig. 2d). Two pairs of grating couplers are required for light at the fundamental and second-harmonic wavelengths separately. The directional couplers are placed between the main waveguide and grating couplers, one on the input side and one on the output side, for multiplexing or demultiplexing light of different wavelengths, which also converts the second-harmonic light between the $TM_{00}$ mode in the input/output waveguide and the $TM_{20}$ mode in the main waveguide.

The devices were characterized with an experimental setup shown in Fig. 2e. First, we needed to calibrate the insertion loss introduced by the grating couplers and directional couplers. From the calibration devices where the grating couplers and directional couplers are connected together (Fig. 2b), the fiber-to-chip loss was found to be ~9.3 dB and ~18.3 dB at the fundamental (telecom) and second-harmonic (near-visible) wavelengths, respectively. Next, by excluding the insertion loss of the grating couplers and directional couplers from measured transmission results of the real devices, we extracted the propagation loss of the main waveguide. By varying the waveguide width *w* near the optimal design parameter 2.48 μm, we obtained the measured and simulated propagation loss of the waveguides at the wavelength of 779.7 nm, as shown in Fig. 2f. The measured results agree well with the simulated results, showing that the propagation loss indeed reduces significantly near the optimal design parameter. These results prove the existence of BIC in the waveguides of the optimal design, which lays the foundation for efficient second-harmonic generation in the subsequent measurements. Figure 2g plots the propagation loss measured from the same waveguides at the wavelength of 1559.4 nm, which is ~2 dB/cm for most of the waveguide devices and does not vary significantly with the waveguide width. This is because the propagation loss of the $TE_{00}$ mode is not subject to the BIC mechanism and is caused by material absorption and surface scattering.

We next measured second-harmonic generation from 5-mm-long waveguide devices. Light at a telecom wavelength from a continuous-wave tunable semiconductor laser (Yenista Tunics-T100S-HP) was amplified by an erbium-doped fiber amplifier (Pritel INHPFA-33-IO), sent through a fiber polarization controller, and then coupled into the device under test as the pump. The generated second-harmonic light was coupled out of the chip and then split equally into two paths. In one path it was collected by a highly sensitive photodetector (Agilent HP 81530A) for recording the power. In the other path, it was sent to an optical spectrum analyzer (Yokogawa AQ6370D) for spectral analysis. By sweeping the wavelength of the pump light and recording the power of the generated second-harmonic light, we obtained the normalized spectrum of second-harmonic conversion efficiency from a fabricated device, as shown in Fig. 3a. The central main peak of the spectrum is located at 1560.9 nm, where the phase-matching condition is perfectly satisfied. Compared with the theoretical conversion efficiency including the effects of phase matching and loss factor in Eq. (1), the measured conversion efficiency spectrum in Fig. 3a has stronger side peaks, which is probably attributed to the nonuniform waveguide geometries along its 5 mm length. The maximal on-chip second-harmonic conversion efficiency measured at the BIC point is 0.175% $W^{-1}$ $cm^{-2}$, which agrees well with the theoretical value (0.20% $W^{-1}$ $cm^{-2}$) when taking into consideration the measured propagation loss of the waveguide at both fundamental and second-harmonic wavelengths.

Figure 3b plots the peak second-harmonic power as a function of the pump power coupled into the on-chip waveguides, which shows an approximately quadratic power dependence of the second-harmonic power on the input pump power as predicted in Eq. (1). The lowest input pump power used in our devices is −4.05 dBm. Figure 3c shows the measured normalized second-harmonic conversion efficiency of the waveguides $P_{\mathrm{SHG}} / P_{\mathrm{pump}}^2 L^2$ as a function of the waveguide width $w$. It is clear that the maximum is reached for the waveguides with the optimal waveguide width $w = 2.48$ μm at the BIC point. As the device structures deviate from the optimal design for the BIC, the second-harmonic generation efficiency drops dramatically. The reduction in the normalized conversion efficiency is attributed mainly to the loss factor $g(L)$ in Eq. (3). According to Eq. (1), the plotted normalized conversion efficiency is actually $g(L)\eta$, which confirms that the second-harmonic generation efficiency depends not only on the theoretical normalized conversion efficiency $\eta$ for an ideal lossless waveguide, but also on the propagation loss $g(L)$ of an actual waveguide. An analysis on the influence of propagation loss is provided in Section S1 in the Supporting Information.

In addition to fabrication easiness, our devices also exhibited excellent thermal stability. Lithium niobate has a positive thermo-optic coefficient[42] ($dn_e/dT \sim 4 \times 10^{-5}$ $K^{-1}$). The phase-matching point and the consequent wavelength for the maximal second-harmonic conversion efficiency have a strong dependence on temperature in devices fabricated with the traditional method[41]. In our devices, the polymer ZEP520A on top of the LNOI substrate has a negative thermo-optic coefficient[43, 44] ($dn/dT \sim -1 \times 10^{-4}$ $K^{-1}$). In this hybrid waveguide structure, since both the fundamental and second-harmonic modes are confined mostly to the lithium niobate layer, the negative thermo-optic coefficient of the polymer can effectively compensate for the change of effective refractive indices of both modes as temperature varies, which makes our devices highly robust against environmental temperature fluctuations. We measured the spectra of second-harmonic conversion efficiency with the device temperature varying from 25°C to 100°C, with the results shown in Figs. 4a and 4b. Figure 4b plots the measured and fitted phase-matched wavelength as a function of temperature. The device shows an excellent thermal stability from 25°C to 80°C where the phase-matched wavelength is varied by only ~0.8 nm. In contrast to devices fabricated with the traditional method[41], the thermal stability of our devices is a clear advantage for many applications such as optical communications, nonlinear conversion, and quantum photonics. In the temperature range of 25°C to 100°C, the measured phase-matched wavelength has a total redshift of 1.7 nm, with a nonlinear temperature dependence which is attributed mainly to the quadratic temperature dependence of the thermo-optic coefficients of lithium niobate[42]. A quadratic fit (red line in Fig. 4b) of the experimental data provides $\lambda = 4.921 \times 10^{-4}(T - 35)^2 + 1565.297$, where $\lambda$ is the phase-matched wavelength in units of nm and $T$ is the temperature in units of °C.

**Conclusion**

In conclusion, we for the first time demonstrated nonlinear optics with photonic bound states in the continuum in etchless lithium niobate nanophotonic waveguides. Specifically, we achieved efficient second-harmonic generation from BIC-based waveguides that could be fabricated with an etchless process on a lithium-niobate-on-insulator platform. By selecting the suitable waveguide modes and optimizing the waveguide geometry, the designed waveguides could satisfy the requirement for achieving the BIC and phase matching simultaneously, which provides strong optical confinement, minimizes the propagation loss, and enables efficient second-harmonic generation. We experimentally demonstrated that the optimally designed waveguides supporting a BIC for the second-harmonic mode provided the most efficient second-harmonic generation, confirming the important role of photonic BICs in nonlinear optics. The temperature-varying measurement showed excellent thermal stability of our devices, which is a clear advantage for various athermal nonlinear applications.

On the BIC-based integrated platform, the second-harmonic conversion efficiency of lithium niobate waveguides can be further improved by reducing the effective modal area and enhancing the spatial overlap between the fundamental and second-harmonic modes in the nonlinear region. This can be achieved by using a material with a higher refractive index for the waveguide, by adopting a LNOI wafer with a thicker lithium niobate layer, or by introducing periodic poling to the lithium niobate thin film (see Section S2 in the Supporting Information). For example, by using a LNOI wafer with a 450-nm-thick *z*-cut lithium niobate layer with periodic poling process, the second-harmonic conversion efficiency can achieve 2571% $W^{-1}$ $cm^{-2}$. With simplified fabrication process and compatibility with different optical functional materials, our experimental demonstration will stimulate further exploration of photonic BICs in nonlinear optics for a broad range of applications including optical parametric generation, signal processing, and quantum photonics.

**Notes**

The authors declare no competing financial interest.

**Acknowledgment**

This work was supported by the Research Grants Council of Hong Kong (14208717, 14206318, 14209519).

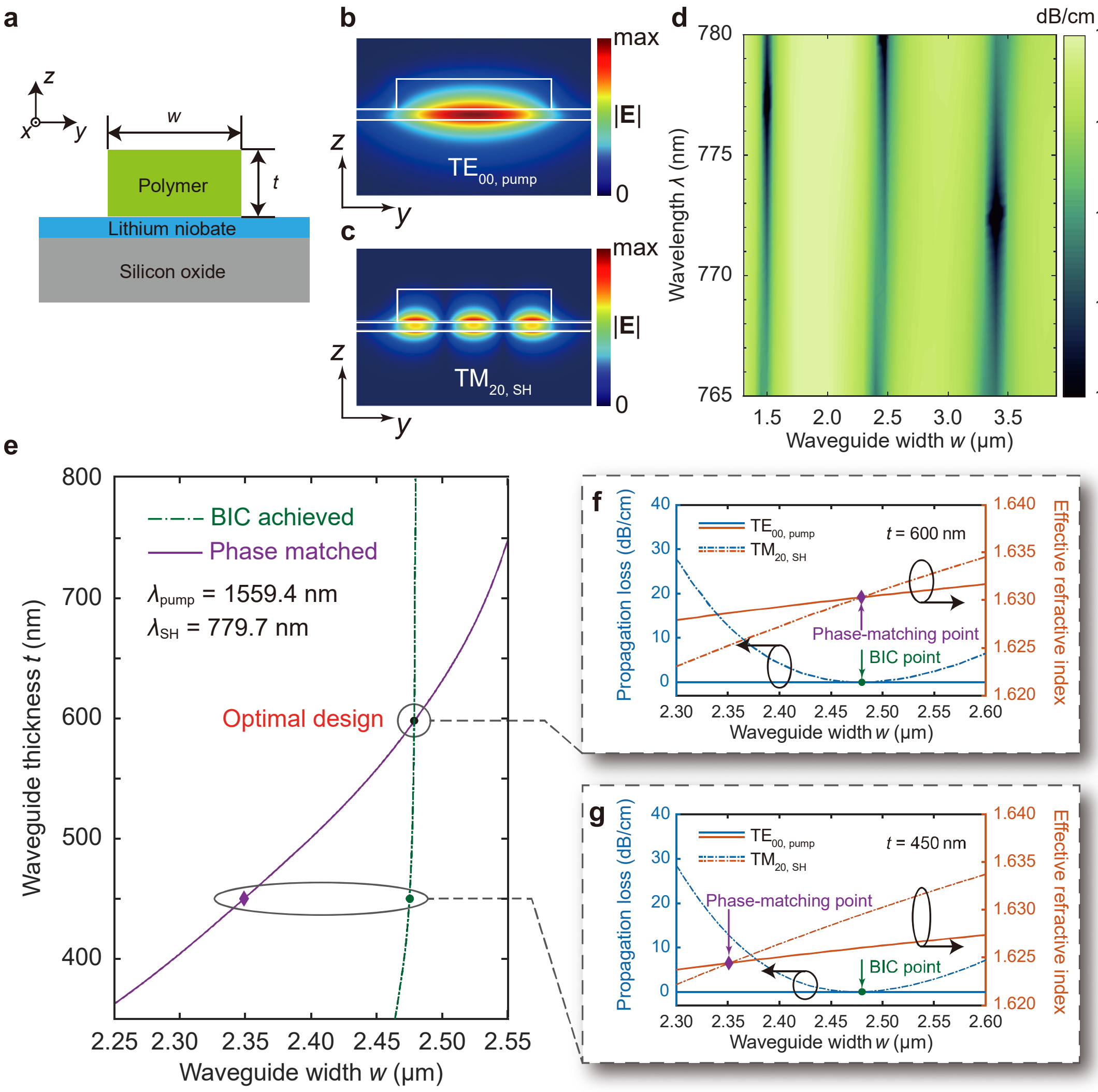

**Figure 1. a** Cross-sectional illustration of the waveguide structure which consists of an easy-to-fabricate polymer waveguide on top of a lithium-niobate-on-insulator substrate. **b**, **c** Electric field (|**E**|) profiles of the $TE_{00}$ mode at pump wavelength (b) and the $TM_{20}$ mode at second-harmonic wavelength (c). **d** Propagation loss of the $TM_{20}$ mode as a function of the waveguide width $w$ and wavelength $\lambda$. The propagation loss approaches zero for certain combinations of parameters (black regions) where the BIC is achieved. **e** Required waveguide width $w$ for achieving the BIC (green dash-dotted line) and phase-matching (purple solid line) conditions, for the structure in (a) with different waveguide thicknesses $t$. **f**, **g** Simulated propagation loss and effective refractive index of the $TE_{00}$ mode and $TM_{20}$ mode at the respective wavelengths as a function of the waveguide width $w$, with the waveguide thickness of $t = 600$ nm (f) and $t = 450$ nm (g).

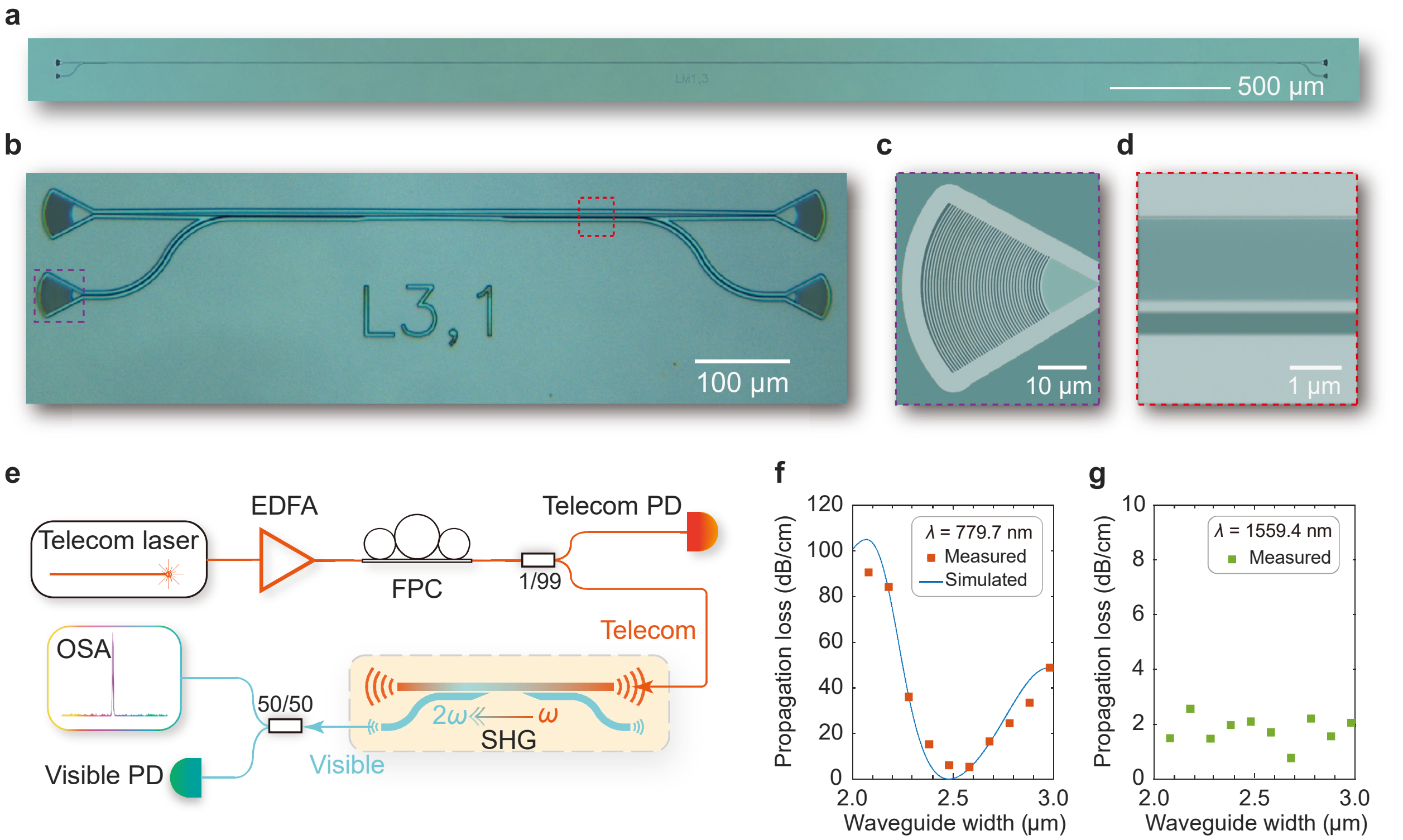

**Figure 2. a** Optical microscope image of an entire fabricated device which consists of a 5-mm-long nonlinear waveguide, directional couplers, and grating couplers. **b** Optical microscope image of a fabricated device for loss calibration which only consists of directional couplers and grating couplers. **c** False-colored scanning electron microscope image of the grating coupler for the second-harmonic light as indicated in (b). **d** False-colored scanning electron microscope image of a part of the directional coupler as indicated in (b). **e** Experimental setup for device characterization and second-harmonic generation measurement. EDFA, erbium-doped fiber amplifier; FPC, fiber polarization controller; PD, photodetector; OSA, optical spectrum analyzer; SHG, second-harmonic generation. **f** Measured and simulated propagation loss of the waveguides at the wavelength of 779.7 nm as a function of the waveguide width $w$. **g** Measured propagation loss of the waveguides at the wavelength of 1559.4 nm as a function of the waveguide width $w$.

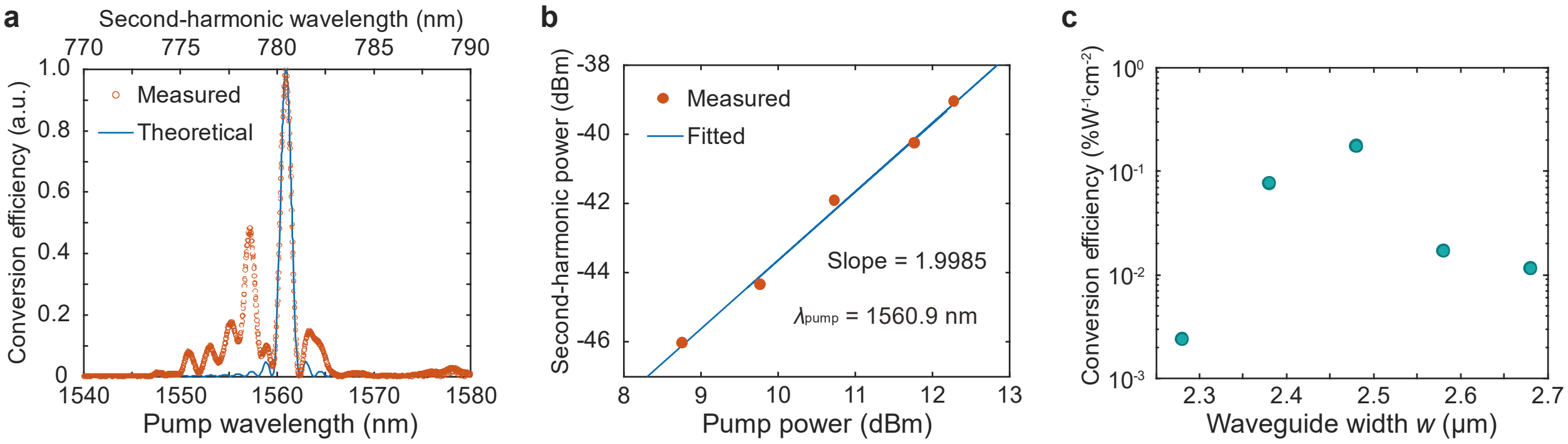

**Figure 3**. **a** Measured (orange circles) and theoretical (blue line) normalized spectra of second-harmonic conversion efficiency of a fabricated device. **b** Measured (orange dots) and fitted (blue line) second-harmonic power as a function of the pump power in the device. **c** Measured second-harmonic conversion efficiency as a function of the waveguide width *w*.

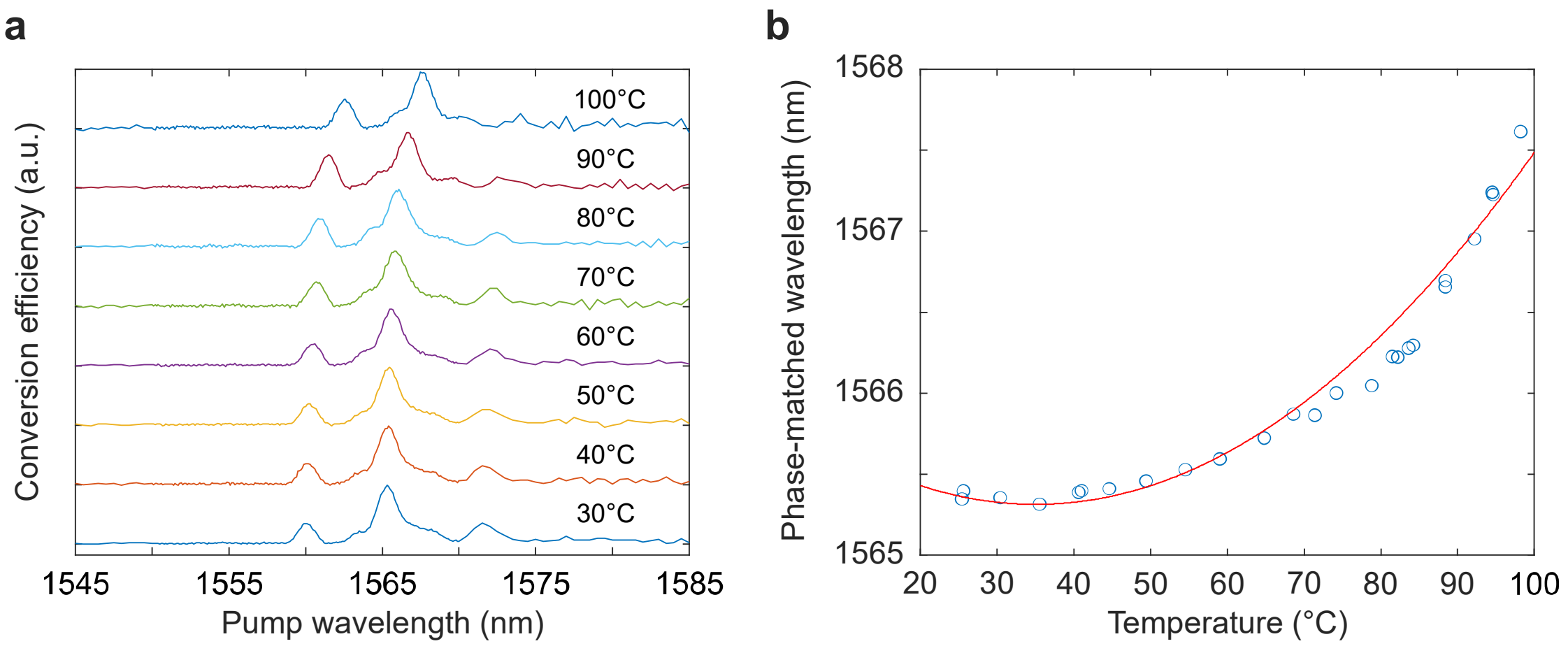

**Figure 4**. **a** Normalized spectra of second-harmonic conversion efficiency of a fabricated device measured at different temperatures. **b** Measured (blue circles) and fitted (red line) phase-matched wavelength as a function of temperature for the device in (a).